\documentclass[aps,prd,twocolumn,groupedaddress,amsfonts,amssymb,amsmath,showpacs,showkeys]{revtex4-1}

\usepackage{graphicx}

\usepackage{psfrag}

\newcommand{\aap}{Astronomy and Astrophysics}

\newcommand{\jcap}{Journal of Cosmology and Astroparticle Physics}

\newcommand{\mnras}{MNRAS}

\newcommand{\physrep}{Physics Reports}

\begin{document}


\title{Recovering MOND from extended metric theories of gravity}

\author{T. Bernal$^{1}$}
\email[Email address: ]{tbernal@astro.unam.mx}
\author{S. Capozziello$^{2,3}$}
\email[Email address: ]{capozzie@na.infn.it}
\author{J.C. Hidalgo$^{1,4}$}
\email[Email address: ]{jhidalgo@astro.unam.mx}
\author{S. Mendoza$^{1}$}
\email[Email address: ]{sergio@astro.unam.mx}
\affiliation{$^1$Instituto de Astronom\'{\i}a, Universidad Nacional
                 Aut\'onoma de M\'exico, AP 70-264, Distrito Federal 04510,
	         M\'exico \\
             $^2$Dipartimento di Scienze Fisiche, Universit\`a degli Studi di
	         Napoli ``Federico II'', Complesso Universitario di Monte
		 Sant'Angelo, Edificio N, via Cinthia, 80126, Napoli, Italy \\
             $^3$I.N.F.N. - Sezione di Napoli, Complesso Universitario di
	         Monte Sant'Angelo, Edificio G, via Cinthia, 80126, Napoli,
	         Italy \\
             \(^4\)Departamento de F\'{\i}sica, Instituto Nacional de
	         Investigaciones Nucleares, AP 18-1027, Distrito Federal
		 11801, M\'exico
            }

\date{\today}

\begin{abstract}
  We show that the Modified Newtonian Dynamics (MOND) regime can be
fully recovered as the weak-field limit of a particular theory
of gravity formulated in the metric approach.  This is possible when
Milgrom's acceleration constant is taken as a fundamental
quantity which couples to the theory in a very consistent manner.  
As  a consequence, the scale invariance of the gravitational
interaction is naturally broken.  In this sense,  Newtonian gravity is
the weak-field limit of general relativity and MOND is
the weak-field limit of that particular extended theory of 
gravity.  We also prove that a Noether's symmetry approach to the problem
yields a conserved quantity coherent with this relativistic MONDian
extension.
\end{abstract}


\maketitle

\section{Introduction}
\label{introduction}

  Milgrom \cite{milgrom83a,milgrom83c} developed a non-relativistic theory
of gravity in order to explain observed flat rotation curves of spiral
galaxies.  This Modified Newtonian Dynamics (MOND) theory  has proved useful in
explaining a great variety of astronomical phenomena without requiring
the presence of a dark matter component (see e.g. \cite{milgrom09,milgrom10}
and references therein).

  As explained by an extended Newtonian approach to gravity,
Mendoza~\textit{et al.}~\cite{mendoza11} showed that the key feature
of MOND is the introduction of a fundamental acceleration \( a_0 \)
in the theory, which on itself at the non-relativistic level, makes
gravity \textit{scale dependent}.

Through the years, finding the relativistic extension of MOND has
become a big challenge.  The most successful attempt was proposed by
Bekenstein \cite{bekenstein04} who formulated a Tensor-Vector-Scalar (TeVeS)
relativistic theory of MOND. This approach requires tensor, vector and
scalar fields to achieve a self-consistent description.  However, the
many cumbersome mathematical complications that TeVeS present are evident.
Furthermore, it cannot reproduce crucial astrophysical phenomena (see
e.g. \cite{sakellariadou09}).

  On the other hand, since extended metric \( f(R) \) theories
of gravity have proved very successful on a wide variety
of cosmological scenarios \citep[see e.g.][and references
therein]{capozziellobook,sotiriou10,nojiri06,nojiri11} it is natural
to seek a relativistic generalisation of MOND in this direction, which
has not deserved too much attention since Soussa \cite{soussa03b}
and Soussa and Woddard \cite{soussa03a} developed a no-go theorem
which prevented all metric \( f(R) \) theories of gravity to become
relativistic candidates for MOND.  However, Mendoza and Rosas-Guevara
\cite{mendoza07} showed counterexamples of this no-go theorem disproving
its general validity.  Furthermore, the works by Capozziello {\it et
al.} \cite{capozziello06,capozziello07a,capozziello09}  and Sobouti
\cite{sobouti06} made clear that particular \( f(R) \) models are
capable of explaining phenomena usually ascribed to MOND.  Up to now,
the challenge has been to formulate a suitable \( f(R) \) theory able
to converge to standard MOND in the non-relativistic regime.

 In this article,  we show how a particular \( f(R) \) metric theory
of gravity, derived from first principles, is capable of reproducing
MOND when its non-relativistic regime is reached.  To do so, in
Section~\ref{dimensional} we set the foundations of a metric theory
of gravity with the use of correct dimensional quantities. Then,
in Section~\ref{order-magnitude} we solve the problem of a point
mass source on a static space-time and find, at first order of
approximation, the particular form of the function \( f(R) \).
In Section~\ref{weak-field} the integration constants of the theory
are fixed by solving the same problem in the formalism of metric
perturbations. In Section~\ref{noether} we show that the existence of
a Noether symmetry confirms the appearance of a characteristic scale of
the problem in the MONDian regime.  Finally in Section~\ref{discussion}
we comment on the obtained results.

\section{Dimensional grounding}
\label{dimensional}

  Let us assume that a point mass \( M \) located at the origin
of coordinates generates a relativistic gravitational field in the MONDian
regime and that a metric formalism describes the field equations.

 This problem is characterised by the following quantities: the
speed of light in vacuum \( c \), the mass \( M \) of the central object
generating the gravitational field, Newton's constant of gravity \( G \)
and Milgrom's acceleration constant \( a_0 \).  With these parameters,
two ``fundamental lengths'' can be built:

\begin{equation}
  r_\text{g} := \frac{ G M }{ c^2 }, \qquad l_M := \left( \frac{ G M }{ a_0
    }\right)^{1/2}.
\label{eq01}
\end{equation}

\noindent The gravitational radius \( r_\text{g} \) is a  length that
appears once relativistic effects are introduced on a theory of gravity.
The mass-length scale \( l_M \), as described in \cite{mendoza11}, is
a characteristic length which appears on a gravitational theory when
MONDian effects are to be taken into account (for consistency we note
here that a third length, $ \lambda := { l_M^2} / { r_\text{g} } = {
c^2 } / { a_0 }$, that does not contain the mass, can be constructed as
a combination of the previous two).

  As discussed by Mendoza~\textit{et al.}~\cite{mendoza11}, in
the non-relativistic regime, a test particle located at the radial
coordinate \( r \) from the origin will obey the MONDian dynamics when
\( l_M /r \ll 1 \).  When \( l_M / r \gg 1 \) the gravitational field
is Newtonian.  As such, when relativistic effects are taken into account
for the gravitational field, then standard general relativity should be
recovered in the limit  \( l_M / r \gg 1 \), and a relativistic version
of MOND should be obtained when \( l_M / r \ll 1 \).  This shows that
the pursue of a complete metric description leads one to consider the
scale-dependence of gravity.

  The length scales presented in equation~\eqref{eq01} must somehow
appear in a relativistic theory of gravity which accepts the fundamental
nature of the constant \( a_0 \).  For example, in the metric formalism,
a generalised Hilbert action \( S_\text{H} \) can be written in the
following way:

\begin{equation}
   S_\text{H}  = - \frac{ c^3 }{ 16 \pi G L_M^2 } \int{ f(\chi) \sqrt{-g}
     \, \mathrm{d}^4x},
\label{eq03}
\end{equation}

\noindent which slightly differs from its traditional form (see
e.g. \cite{capozziellobook,sotiriou10,capozziello10a})

\begin{equation}
  S_\text{H} = - \frac{ c^3 }{ 16 \pi G } 
    \int f(R)\sqrt{-g} \, \mathrm{d}^4x\,,
\label{fRaction}
\end{equation}

\noindent  since we have introduced the following dimensionless quantity:

\begin{equation}
  \chi := L_M^2 R,
\label{eq04}
\end{equation}

\noindent where \( R \) is Ricci's scalar and \( L_M \) defines a length
fixed by the parameters of the theory.  The explicit form of the length
\( L_M \) has to be obtained once a certain known limit of the theory is
taken, usually a non-relativistic limit.  Note that the definition of \(
\chi \) gives a correct dimensional character to the action~\eqref{eq03},
something that is not completely clear in all previous works dealing
with a metric description of the gravitational field.  For \( f(\chi)
= \chi \) the standard Einstein-Hilbert action is obtained.

  On the other hand, the matter action has its usual form,

\begin{equation}
  S_\text{m} = - \frac{ 1 }{ 2 c } \int{ {\cal L}_\text{m} \, \sqrt{-g} \,
    \mathrm{d}^4x },
\label{eq05}
\end{equation}

\noindent with \( {\cal L}_\text{m} \) the Lagrangian density of the
system.  The null variations of the complete action, i.e. \( \delta
\left( S_\text{H} + S_\text{m} \right) = 0 \), yield the following
field equations:

\begin{equation}
  \begin{split}
    f'(\chi) \, \chi_{\mu\nu} - \frac{ 1 }{ 2 } f(\chi) g_{\mu\nu} - L_M^2 &
      \left( \nabla_\mu \nabla_\nu -g_{\mu\nu} \Delta \right) f'(\chi)
  				\\
    &= \frac{ 8 \pi G L_M^2 }{ c^4} T_{\mu\nu},
  \end{split}
\label{eq06}
\end{equation}

\noindent where the dimensionless Ricci tensor \( \chi_{\mu\nu} \) 
is given by:

\begin{equation}
  \chi_{\mu\nu} := L_M^2 R_{\mu\nu},
\label{eq07}
\end{equation}

\noindent and \( R _{\mu\nu} \) is the standard Ricci tensor.
The Laplace-Beltrami operator has been written as \( \Delta :=
\nabla^\alpha \nabla_\alpha \) and the prime denotes derivative with
respect to its argument.  The energy-momentum tensor \( T_{\mu\nu} \)
is defined through the following standard relation: \( \delta S_\text{m}
= - \left( 1 / 2 c \right) T_{\alpha\beta} \, \delta g^{\alpha\beta} \).
In here and in what follows, we choose a (\(+,-,-,-\)) signature for
the metric \( g_{\mu\nu} \) and use Einstein's summation convention over
repeated indices.

  The trace of equation~\eqref{eq06} is:
\begin{equation}
  f'(\chi) \, \chi  - 2 f(\chi) + 3 L_M^2  \, \Delta  f'(\chi) = 
    \frac{ 8 \pi G L_M^2 }{ c^4} T,
\label{eq08}
\end{equation}

\noindent where \( T := T^\alpha_\alpha \).

Since we are only interested on the gravitational field produced by
a point mass source located at the origin, then the mass density \(
\rho \) is given by

\begin{equation}
  \rho = M \, \delta(\boldsymbol{r}),
\label{eq09}
\end{equation}

\noindent where \( \delta(\boldsymbol{r}) \) represents the three
dimensional Dirac delta function.  With this, it follows that the 
only non-zero component of the energy-momentum tensor is given by

\begin{equation}
  T_{00} = \rho c^2 = c^2 M \, \delta(\boldsymbol{r}).
\label{eq10}
\end{equation}

  A point mass distribution generates a stationary spherically symmetric
space-time and so, the trace equation~\eqref{eq08} contains all the
relevant information relating the field equations.
In what follows we assume a power~law form for the function \( f(\chi)\),
i.e. 

\begin{equation}
  f(\chi) = \chi^b.
\label{eq11}
\end{equation}

\section{Order of magnitude approach}
\label{order-magnitude}

  Let us first analyse the problem described in the previous Section 
by performing  an order of magnitude
approximation of the trace equation~\eqref{eq08}.  Under these
circumstances, \( \mathrm{d} / \mathrm{d} \chi \approx 1 / \chi \), \(
\Delta \approx - 1 / r^2  \) and the mass density \( \rho \approx  M /
r^3 \).  This approximation implies that equation~\eqref{eq08} takes the
following form:

\begin{equation}
   \chi^b  \left( b - 2 \right) - 3 b L_M^2  \frac{ \chi^{(b-1)} }{ r^2 }
     \approx \frac{ 8 \pi G M L_M^2 }{ c^2 r^3}.
\label{eq12}
\end{equation}

Note that the second term on the left-hand side of equation~\eqref{eq12}
is much greater than the first term when the following condition is
satisfied:

\begin{equation}
  R  r^2  \lesssim \frac{ 3 b }{ 2 - b }.
\label{eq13}
\end{equation}

\noindent  At the same order of approximation, Ricci's scalar \( R
\approx \kappa = R_\text{c}^{-2} \), where \( \kappa \) is the Gaussian
curvature of space and \( R_\text{c} \) its radius of curvature and so,
relation~\eqref{eq13} essentially means that

\begin{equation}
  R_c \gg r.
\label{eq14}
\end{equation}

\noindent  In other words, the second term on the left-hand side of
equation~\eqref{eq12} dominates the first one when the local radius of
curvature of space is much grater than the characteristic length \(
r \).  This should occur in the weak-field regime, where MONDian effects
are expected.  For a metric description of gravity, this limit must
correspond to the relativistic regime of MOND. In this article  we will
only deal with this approximation.  At the end of the current Section
we show an equivalent relation to inequality~\eqref{eq14} which has a
more physical meaning.

  Under assumption~\eqref{eq14}, equation~\eqref{eq12} takes the following
form:

\begin{equation}
    R^{ (b-1)} \approx - \frac{ 8 \pi G M  }{ 3 b c^2 r
      L_M^{2 \left( b - 1 \right) } }.
\label{eq15}
\end{equation}

  We now recall the well known relation followed by the Ricci scalar at
second order of approximation at the non-relativistic level 
\cite{daufields}:
\begin{equation}
  R =  - \frac{ 2 }{  c^2 } \nabla^2 \phi = +\frac{ 2 }{ c^2 } \nabla \cdot
    \boldsymbol{a},
\label{eq16}
\end{equation}

\noindent where the negative gradients of the gravitational potential \(
\phi \) provide the acceleration \( \boldsymbol{a} := - \nabla \phi \)
felt by a test particle on a non-relativistic gravitational field. At
order of magnitude, equation~\eqref{eq16} can be approximated as

\begin{equation}
  R \approx - \frac{ 2 \phi }{  c^2 r^2 } \approx  \frac{ 2 a }{  c^2 r }.
\label{eq17} 
\end{equation}

  Substitution of this last equation on relation~\eqref{eq15} gives

\begin{eqnarray}
  a &\approx&   - \frac{ c^2 r }{ 2 L_M^2 }  \left( \frac{ 8 \pi G M  }{
    3 b c^2 r  } \right)^{1/\left( b - 1 \right)}\nonumber\\ & \approx
    & - c^{\left( 2 b- 4  \right)/\left( b - 1 \right) } r^{ \left( b -
    2 \right) / \left( b - 1 \right) } L_M^{-2} \left( G M  \right)^{
    1 / \left( b - 1 \right) }.
\label{eq19}
\end{eqnarray}

  This last equation converges to a  MOND-like acceleration \( a \propto
1 / r \) if \( b - 2 = - \left( b - 1  \right) \), i.e. when

\begin{equation}
  b = 3 / 2.  
\label{eq20} 
\end{equation}

\noindent Also, at the lowest order of approximation, in the extreme
non-relativistic limit, the velocity of light \( c \) should not appear
on equation~\eqref{eq19} and so, the only possibility is that \( L_M \)
depends on a power of \( c \), i.e.

\begin{equation}
  L_M^{ - 2 } \propto c^{\left( 4 - 2 b \right)/\left( b - 1 \right) }  =
    c^2, \quad \text{and so,} \qquad L_M \propto c^{-1}.
\label{eq21}
\end{equation}

  As discussed in Section~\ref{dimensional}, the length \( L_M \) must
be constructed by fundamental parameters describing the theory of 
gravity and so, let us
assume that 

\begin{equation}
  L_M = \zeta \, r_\text{g}^\alpha l_M^\beta, \qquad \text{with} \qquad
    \alpha + \beta = 1,
\label{eqlm}
\end{equation}

\noindent where the constant of proportionality \( \zeta \) is a
dimensionless number of order one that will be formally obtained in
Section~\ref{weak-field}. Substituting equation~\eqref{eqlm}  and the
value obtained in~\eqref{eq20} into relation~\eqref{eq21}
it then follows that

\begin{equation}
  \alpha = \beta = 1/2, \qquad \text{i.e.} \qquad L_M \approx
    r_\text{g}^{1/2} l_M^{1/2} \, .
\label{eq22}
\end{equation}

  If we now substitute this last result and relation~\eqref{eq20} 
in equation~\eqref{eq19} it follows that

\begin{equation}
  a \approx - \frac{ \left( a_0 G M \right)^{1/2} }{ r },
\label{eq23}
\end{equation}

\noindent which is the traditional form of MOND in spherical symmetry (see
e.g.  \cite{milgrom09,milgrom10,bekenstein06} and references therein).
Also, the results of equation~\eqref{eq23} in~\eqref{eq17} mean that

\begin{equation}
  R \approx \frac{ r_\text{g} }{ l_M } \, \frac{ 1 }{ r^2 },
\label{eq24}
\end{equation}

\noindent and so, inequality~\eqref{eq14} is equivalent to

\begin{equation}
  l_M \gg r_\text{g}.
\label{eq25}
\end{equation}

\noindent  The regime imposed by equation~\eqref{eq25} is precisely the
one for which MONDian effects should appear in a relativistic theory of
gravity.  This is an expected generalisation of the results presented by
Mendoza~\textit{et al.}~\cite{mendoza11} in the weak field limit regime for
which \( l_M \ll r \) and so, combining this with equation~\eqref{eq25}
yields \( r \gg l_M \gg r_\text{g} \).  In this connection, we also note
that Newton's theory of gravity is recovered in the limit \( l_M \gg r \gg
r_\text{g} \).

\section{Weak field limit approach}
\label{weak-field}

 We now use the trace~\eqref{eq08} to the lowest order of
perturbation.  Results of this perturbation in the Newtonian
limit for other metric theories of gravity have been reported by
\cite{capozziello07,capozziello10}.  However, since we are interested
in the lowest order of approximation in the MONDian regime, we expect
different results.

  Under the assumption of spherical symmetry for a static space-time,
with a diagonal metric given by

\begin{align}
  g_{00} &= 1 + \frac{ 2 \phi }{ c^2 },  &\quad g_{11} &= -1, \nonumber \\
  g_{22} &= -r^2, &\quad g_{33} &= -r^2 \sin^2{\theta},
\label{eq27z}
\end{align}

\noindent then

\begin{equation}
  \begin{split}
   \Delta f'(\chi) &= \frac{ 1 }{ \sqrt{-g} } \frac{ \partial }{ \partial
      x^\mu } \left( \sqrt{-g} \, g^{\mu\nu} \frac{ \partial f'(\chi)}{
      \partial x^\nu } \right),
    				\\
    &= - \frac{ 1 }{ r^2 } \frac{ \partial }{ \partial r
    } \left( r^2 \frac{ \partial f'(\chi) }{ \partial r } \right) = -
    \nabla^2 f'(\chi), 
  \end{split}
\label{eq27}
\end{equation}

\noindent at the lowest order of perturbation in \( \phi / c^2 \). In this
case, when condition~\eqref{eq14} or equivalently equation~\eqref{eq25} is
satisfied, the trace~\eqref{eq08} of the field equations is given by:

\begin{equation}
  - 3 \nabla^2 f'(\chi) = \frac{ 8 \pi G
  }{ c^2 } \rho.
\label{eq28}
\end{equation}

\noindent Note that this equation can also be obtained by performing
a direct perturbation of \eqref{eq08}.  This is so because in the weak
field limit, the field equations~\eqref{eq01} are studied at orders of
powers of \( c^{-2} \) \cite{capozziello07,capozziello09}. At the lowest
zeroth-order of the perturbation, the Ricci scalar \( ^{(0)}R = 0 \)
everywhere and so, it describes a flat space-time.  At the next second
perturbation order $ \mathcal{O}(\chi) = \mathcal{O}(L_{\mathrm M}^2) +
\mathcal{O}( ^{(2)} R ) $, where \( ^{(2)}R \) represents Ricci's scalar
at the second order of the perturbation.  Since \( \mathcal{O}( L_M ) =
1 \) according to relation~\eqref{eq21}, then \( \mathcal{O}(\chi) = 4 \).
If we now use the power law relation~\eqref{eq11} with~\eqref{eq20} and
perturb the trace~\eqref{eq08}, it follows that the first two terms on
the right-hand side are of order $ \mathcal{O}(\chi^{3/2}) = 6 $, while
the remaining two terms are of order \( \mathcal{O}( L_M^2 \chi^{1/2}
) = 4 \) and so, it follows that the trace~\eqref{eq08} at its lowest
non-zero perturbation order is exactly relation~\eqref{eq28}.

  Equation~\eqref{eq28} can be integrated straightforward using the standard
Poisson equation of Newtonian gravity:

\begin{equation}
  \nabla^2 \phi_\text{N} = 4 \pi G \rho,
\label{eq28a}
\end{equation}

\noindent for the Newtonian potential \( \phi_\text{N} \).  Substitution of
this relation on~\eqref{eq28} yields

\begin{equation}
  \nabla^2 \left( f'(\chi) + \frac{ 2 }{ 3 c^2 } \phi_\text{N} \right) = 0.
\label{eq28b}
\end{equation}

\noindent  The trivial solution of the previous Laplace-like equation
occurs when the argument of the Laplacian is equal to zero, and so

\begin{equation}
  f'(\chi) = - \frac{ 2 }{ 3 c^2 } \phi_\text{N}.
\label{eq28c}
\end{equation}

  Substitution of 
equations~\eqref{eq11},~\eqref{eq20},~\eqref{eqlm}, and \eqref{eq22} in
relation~\eqref{eq28c} leads to

\begin{equation}
  R = \left( \frac{ 4 }{ 9 } \right)^2 \zeta^{-2} \left( \frac{ r_\text{g}
    }{ l_M } \right) \frac{ 1 }{ r^2 },
\label{eq29}
\end{equation}

\noindent \noindent where we have used the fact that 
for a point mass \( M \), the Newtonian potential is given by: 

\begin{equation}
  \phi_\text{N} = - G \frac{ M }{ r }.
\label{eq30}
\end{equation}

\noindent We now substitute relation~\eqref{eq16} into equation~\eqref{eq29} to
obtain:

\begin{equation}
  \frac{ 2 }{ c^2 } \nabla \cdot \boldsymbol{a} = \left( \frac{ 4 }{
    9 } \right)^2 \zeta^{-2} \left( \frac{ r_\text{g} }{ l_M } \right)
    \frac{ 1 }{ r^2 }.
\label{eq31}
\end{equation}
 
  Integrating this equation over a spherical volume of radius \( r \)
and using Gauss's theorem on the left-hand side, we obtain

\begin{equation}
  a = - \left( \frac{ 2 \sqrt{2} }{ 9 \zeta } \right)^2 \left( \frac{
    c^2 r_\text{g} }{ l_M } \right) \frac{ 1 }{ r },
\label{eq32}
\end{equation}

\noindent and so, by choosing

\begin{equation}
  \zeta = \frac{ 2 \sqrt{2} }{ 9 },
\label{eq33}
\end{equation}

\noindent we reach the MOND acceleration limit

\begin{equation}
  a = - \frac{ \left( a_0 G M \right)^{1/2} }{ r },
\label{eq34}
\end{equation}

\noindent at the lowest perturbation order of the theory. 

  The formulation of the theory described so far implies that, as soon as
we relax the hypothesis for which the gravitational action is strictly
the one described by standard general relativity, new characteristic
lengths have to be considered. In the following Section, we show that
such scales coherently appear as constants of motion of the problem.

\section{Noether's symmetries}
\label{noether}

  The above results are based on the fact that we are assuming the
power~law relation~\eqref{eq11} for the gravitational action. In
particular, for \( f(R) \propto R^{3/2} \) the MOND acceleration regime
is recovered. This particular theory also admits exact cosmological
solutions where a matter dominated era evolves towards the accelerated
universe observed today \cite{prado} and recovers the observed dynamics at
astrophysical scales \cite{rubano}. Also, this solution is particularly
relevant since its conformal transformation is exactly invertible, as
shown in \cite{f(R)-cosmo}.  Noether's symmetries give rise to conserved
quantities that are directly related to characteristic length scales
\cite{prado}.

  In order to develop a Noether's approach to the problem, let us consider
$f(\chi)$ gravity in static spherical symmetry, following the same
ideas as the ones exposed in Section \ref{weak-field}. The spherical 
point--like $f(\chi)$ Lagrangian can be obtained by imposing the 
spherical symmetry directly into the action \eqref{eq03}. As a 
consequence, the infinite
number of degrees of freedom of the original field theory will be
reduced to a finite number. The technique is based on the choice of a
suitable Lagrange multiplier defined by assuming the known explicit form of 
Ricci's scalar $R$ \cite{noether}.

  The static spherically symmetric metric can be expressed as

\begin{equation}
  \mathrm{d}s^2 = A(r) c^2 \mathrm{d}t^2 - B(r) \mathrm{d}r^2 - C(r)
    \mathrm{d}\Omega^2,
\label{me2}
\end{equation}

\noindent where $ \mathrm{d}\Omega^2 := \mathrm{d}\theta^2 +
\sin^2{\theta} \mathrm{d}\varphi^2 $ is the angular displacement and \(
C(r) := r^2 \).  Since we are interested in the weak-field limit approach
of the $f(\chi)$ theory, let us assume $B(r)=1$.

  The point--like $f(\chi)$ Lagrangian \( L \) is obtained by rewriting
the action~\eqref{eq03} as

\begin{equation}
  S = - \frac{ c^3 }{ 16 \pi G L_M^2 } \int{ \left[ f(\chi) -
      \lambda (\chi - \bar{\chi}) \right] \sqrt{-g} \, \mathrm{d}^4x},
\label{lma}
\end{equation}

\noindent where $\lambda$ is a Lagrange multiplier and
$\bar{\chi} = L_M^2 \bar{R}$, for the known Ricci scalar $\bar{R}$
expressed in terms of the metric~\eqref{me2} with $B(r)=1$:

\begin{equation}
  \bar{R} = \frac{A''}{A} + \frac{2 C''}{C} + \frac{A'C'}{AC} -
            \frac{{A'}^2}{2 A^2} - \frac{{C'}^2}{2 C^2} - \frac{2}{C}.
\label{ricci}
\end{equation}

\noindent In the previous equation, the prime denotes the derivative with
respect to the radial coordinate $r$.   Variations of the action with
respect to \( \chi \) give the explicit form of the Lagrange multiplier:

\begin{displaymath}
  \lambda = \frac{ \mathrm{d} f(\chi) }{ \mathrm{d} \chi } := f_\chi.
\end{displaymath}

  Substituting this result in the action~\eqref{lma} and eliminating the
boundary terms (cf. \cite{noether}), the point--like Lagrangian is
obtained:

\begin{equation}
  \begin{split}
    \mathrm{L} = & - \frac{L_M^2}{\sqrt{A}} \left[ \frac{A
      f_\chi}{2C}{C'}^2 + f_\chi A' C' + C f_{\chi\chi}A' \chi' +
      \right. \\
    & 2 A f_{\chi\chi} C' \chi' \bigg] - \sqrt{A} \left[
      (2 L_M^2 + C \chi) f_\chi - C f \right].
  \end{split}
\label{lag1}
\end{equation}

\noindent Note that this Lagrangian is canonical since only the generalised
positions, $ \boldsymbol{q} = (A,C,\chi) $, and their generalised
velocities $ \boldsymbol{q}'= (A',C',\chi') $, appear explicitly.

  In the MONDian regime, where equation~\eqref{eq14} holds, the last two
terms on the right-hand side of equation~\eqref{lag1} are of order \( Cf
\approx C \chi^{3/2} \) and so, both are much smaller than \( L_M^2 f_\chi \).
This statement is also true at the lowest non-zero order of perturbation,
since all terms on the right-hand side of equation~\eqref{lag1} are of
order 4, except the last two which are of order 6.  With this, it follows
that in the MONDian regime the Lagrangian~\eqref{lag1} can be written as

\begin{equation}
  \begin{split}
    \mathrm{L} = & - \frac{L_M^2}{\sqrt{A}} 
      \left[ \frac{A f_\chi}{2C}{C'}^2 + f_\chi A' C' + 
      C f_{\chi\chi}A' \chi' + \right. \\
    & 2 A f_{\chi\chi} C' \chi' + 2 A \bigg].
  \end{split}
\label{lag2}
\end{equation}

  We now search for symmetries related to the cyclic variables and then
reduce the dynamics. According to Noether's theorem, the existence of
symmetry properties for the dynamics described by the Lagrangian implies
the existence of conserved quantities \cite{arnold,marmo,morandi}. In
principle, this approach allows to select particular $f(\chi)$ gravity
models compatible with spherical symmetry.

  A conserved quantity exists if the Lie derivative of the Lagrangian
\eqref{lag2} along the vector field \textbf{X} vanishes:

\begin{equation}
  \mathcal{L}_{\boldsymbol{\mathrm{X}}} \mathrm{L} = \alpha_i \nabla_{q_i}
  \mathrm{L} + \alpha'_i \nabla_{q'_i} \mathrm{L} = 0,
\label{lie}
\end{equation}

\noindent for $\, i=1,2,3,$ in the configuration space $(A,C,\chi)$.
Solving equation~\eqref{lie} means to find out the functions
$\alpha_{i}$ which constitute the Noether vector \cite{arnold,marmo}.
However, the relation~\eqref{lie} implicitly depends on the form of
$f(\chi)$ and then, by solving it, we also get $f(\chi)$ models
compatible with spherical symmetry. On the other hand, by fixing
the $f(\chi)$ function, we can explicitly solve~\eqref{lie}.
In principle, the same procedure can be worked out any time Noether's
symmetries are identified \cite{defelice}.

  The general form of the Noether vector is given by the 
solution of the Killing equations for the components \( \alpha_i \) of the 
vector \( \boldsymbol{\alpha} \) in a flat space-time (cf. \cite{townsend}):

\begin{gather}
  \alpha_1 = k_1 A + p_1, \nonumber \\
  \alpha_2 = k_2 C + p_2, \nonumber \\
  \alpha_3 = k_3 \chi + p_3,
\label{alpha}
\end{gather}

\noindent with $k_i$ and $p_i$ constants. Using the power~law~\eqref{eq11}
and the general solution~\eqref{alpha} in equation~\eqref{lie}, 
it follows that 

\begin{equation}
  \boldsymbol{\alpha} = \biggl(2(1-b)kA,\ 0,\ k\chi \biggr) \, ,
\label{}
\end{equation}

\noindent where $k$ is an integration constant and we have assumed in the
calculations that $b \neq 1$, $\chi' \neq 0$ and $2AC' + A'C \neq 0$ in
order to obtain solutions different from general relativity (cf. Section
\ref{order-magnitude}).  For this case, the related constant of motion,
$\Sigma_{0}$, is given by

\begin{equation}
  \begin{split}
    \Sigma_{0} &=  \alpha_i \nabla_{q_i'}{\mathrm L}, 
    					\\
    &= L_M^2 b (b-1) k A^{-1/2} C \chi^{b-2} \left[2 (b-1) A \chi' 
    - A' \chi \right]. 
  \end{split}
\label{cm}
\end{equation}

  In general relativity, where \( b = 1 \), Noether's symmetry
approach gives \( \Sigma_0 = 2 r_\text{g} \), which is exactly
the Schwarzschild radius (cf. \cite{townsend}).  On the other hand, in
the MONDian regime, where equation~\eqref{cm} is valid for \( b = 3/2
\), \( C(r) = r^2 \) and at the lowest order of perturbation $A(r)=1 +
2\phi/c^2 $, the related constant of motion is given by

\begin{equation}
  \Sigma_0 = \frac{3}{2} k r_\text{g}^2 l_M.
\label{sigma-mond}
\end{equation}

  The key point in this relation is that it includes two
characteristic lengths, namely the mass length-scale \( l_M \) and the
gravitational radius \( r_\text{g} \).  This is coherent with the results
discussed on the previous Sections, since both characteristic lengths must
appear on a correct relativistic metric theory of MOND.

\section{Discussion}
\label{discussion}

  We have shown  that a metric theory of gravity \( f( \chi ) =
\chi^{3/2} \), where the dimensionless Ricci scalar \( \chi \) is given by
equation~\eqref{eq04}, converges to MOND in the non-relativistic regime.
We note that a previous attempt in this direction was made in the B.Sc.
thesis of Rosas-Guevara \cite{rosas06}, where a metric \( f(R) \propto
R^{3/2} \)  theory was proposed to account for relativistic effects
of MOND.  However, the correct approximation~\eqref{eq25} was never
introduced on that analysis and therefore, the MONDian limit was not
achieved on that work.

  We have also shown that the appearance of a new characteristic length
related to MOND's acceleration is coherent with Noether's symmetries
related to the problem.

  The metric theory of gravity presented here is by no means a complete
description at all scales of gravitation.  It only deals with the MONDian
regime of gravity, i.e.  when equation~\eqref{eq25} is valid. In other
words, our description breaks the scale invariance of gravity in a more
general way than the one described in \cite{mendoza11}.

The mass dependence of \( \chi \) means that the mass needs to appear
on Hilbert's action~\eqref{eq03}.  This is traditionally not the case,
since that action is thought to be purely a function of the geometry
of space-time due to the presence of mass and energy.  However, it
was Sobouti \cite{sobouti06} who first encountered this peculiarity in
the  Hilbert action when dealing with a metric generalisation of MOND.
Following his remarks \cite{sobouti06}, one should not be surprised if
some of the commonly accepted notions, even at the fundamental level of
the action, require generalisations and re-thinking.  An extended metric
theory of gravity goes beyond the traditional general relativity ideas
and in this way, we probably need to change our standard view of its
fundamental principles.

\section{Acknowledgements}

  We thank Xavier Hernandez, Cosimo Stornaiolo, Luis A. Torres and Diego A.
Carranza for the deep comments made to this work.  We are also grateful to
an anonymous referee for his suggestions to the first draft of the article.
This work was supported by a DGAPA-UNAM grant (PAPIIT IN116210-3).
The authors TB, JCH \& SM  acknowledge economic support from CONACyT:
207529, 51009, 26344. JCH acknowledges financial support from CTIC-UNAM.
SC is grateful for the hospitality of IA-UNAM, where the key ideas of
this article first originated.

\bibliographystyle{aipnum4-1long}
\bibliography{fR_MOND}

\end{document}